\documentclass[aip,jcp,reprint]{revtex4-1}
\usepackage{mathtools}
\usepackage{amsmath}
\usepackage{multirow}
\usepackage[english]{babel}
\usepackage{amsmath,lipsum}
\usepackage{etoolbox}
\newcommand{\zerodisplayskips}{%
  \setlength{\abovedisplayskip}{5pt}%
  \setlength{\belowdisplayskip}{5pt}%
  \setlength{\abovedisplayshortskip}{5pt}%
  \setlength{\belowdisplayshortskip}{5pt}}
\appto{\normalsize}{\zerodisplayskips}
\appto{\small}{\zerodisplayskips}
\appto{\footnotesize}{\zerodisplayskips}
\usepackage{graphicx}
\usepackage{color}
\newcommand{\pt}{\textcolor{black}} 

\usepackage{ulem}

\renewcommand\thesubsection{\arabic{subsection}}
\renewcommand\thesubsubsection{\arabic{subsubsection}}
\makeatletter
    \def\@seccntformat#1{\@ifundefined{#1@cntformat}%
       {\csname the#1\endcsname\space}
       {\csname #1@cntformat\endcsname}}
    \def\subsection@cntformat{\thesection.\thesubsection\space} 
    \def\subsubsection@cntformat{\thesection.\thesubsection.\thesubsubsection\space}
\makeatother

\DeclareUnicodeCharacter{0301}{\'{e}} 

\begin{document}

\title{SGOOP-d: Estimating kinetic distances and reaction coordinate dimensionality for rare event systems from biased/unbiased simulations}

\author{Sun-Ting Tsai}
\affiliation{Department of Physics and Institute for Physical Science and Technology, University of Maryland, College Park 20742, USA.}

\author{Zachary Smith}
 \affiliation{Biophysics Program and Institute for Physical Science and Technology,
 University of Maryland, College Park 20742, USA.}

\author{Pratyush Tiwary*}
\email{ptiwary@umd.edu}
\affiliation{Department of Chemistry and Biochemistry and Institute for Physical Science and Technology,
 University of Maryland, College Park 20742, USA.}

\date{\today}

\begin{abstract}

Understanding kinetics including reaction pathways and associated transition rates is an important yet difficult problem in numerous chemical and biological systems especially in situations with multiple competing pathways. When these high-dimensional systems are projected on low-dimensional coordinates, which are often needed for enhanced sampling or for interpretation of simulations and experiments, one can end up losing the kinetic connectivity of the underlying high-dimensional landscape.  Thus in the low-dimensional projection metastable states might appear closer or further than they actually are. To deal with this issue, in this work we develop a formalism that learns a multi-dimensional yet minimally complex reaction coordinate (RC) for generic high-dimensional systems. When projected along this RC, all possible kinetically relevant pathways can be demarcated and the true high-dimensional connectivity is maintained. One of the defining attributes of our method lies in that it can work on long unbiased simulations as well as biased simulations often needed for rare event systems. We demonstrate the utility of the method by studying a range of model systems including conformational transitions in a small peptide Ace-Ala$_3$-Nme, where we show how two-dimensional and three-dimensional reaction coordinate found by our previously published spectral gap optimization method ``SGOOP" [P. Tiwary and B. J. Berne, Proc. Natl. Acad. Sci. 113, 2839 (2016)] can capture the kinetics for 23 and all 28 out of the 28 dominant state-to-state transitions respectively.
\end{abstract}

\maketitle

\section{Introduction}
\label{sec:introduction}
It has been a problem of longstanding theoretical and practical interest to model reaction pathways and transition mechanisms in generic chemical and biological systems.\cite{juraszek2013efficient,tiwary2017and,niu2018molecular,tsai2019reaction,lee2016multiple,roca2018monovalent,prinz2011markov,nagel2020msmpathfinder} Due to recent progress in high-performance computing, brute-force Molecular Dynamics (MD) simulations with all-atom resolution have enabled a possible way to do such analysis in femtosecond temporal and all-atom spatial precision, making it a useful tool for studying diverse phenomena. However, this leads to a deluge of data resulting from explicit enumeration of all atomic coordinates over a very large number of MD timesteps. To make sense of such high-dimensional trajectories resulting from MD, it is a common practice to project them along low-dimensional coordinates identified with one of many dimensionality reduction schemes.\cite{piana2012protein,best2005reaction,hummer2015optimal,tribello2019using} However, more often than not in such schemes, one ends up losing the kinetic connectivity of the high-dimensional landscape. This can thus lead to incorrect interpretation of MD trajectories, for example making molecular conformations appear closer to each other than they are and obfuscating interconversion pathways between them.\cite{altis2008construction} \\

In this work, we develop a formalism that learns a multi-dimensional yet minimally complex reaction coordinate (RC), such that when projected along this RC, all possible kinetically relevant pathways can be demarcated and the true high-dimensional connectivity is maintained. The central idea is to calculate the interconversion times between different pairs of metastable states, which can be defined \textit{a priori} or learned on-the-fly,\cite{smith2018multi} and monitor how these distances change by adding additional dimensions to the RC. The procedure is stopped when the interconversion times do not vary with additional RC components. The interconversion times are calculated using the commute distance framework proposed by No\'{e}, Clementi, and co-workers.\cite{noe2015kinetic,noe2016commute} While such a kinetic or commute distance-based procedure is indeed already recommended best practice in the construction of Markov State Models (MSMs),\cite{husic2018markov} it is not directly amenable to rare event systems that might be undersampled, or accessible only through biased simulations. \\

To deal with this issue, in this work we combine the commute distance\cite{noe2016commute,noe2015kinetic} with the Maximum Caliber based ``Spectral Gap Optimization of Order Parameters (SGOOP)" approach.\cite{tiwary2016spectral} This amounts to inducing a distance metric, which we call ``SGOOP-d" that preserves  kinetic truthfulness, and can be calculated from long unbiased simulations as well as biased simulations. Such biased simulations are often unavoidable in the study of rare events in chemical and biological physics. Here we use metadynamics\cite{valsson2016enhancing} as an example of the biasing method to illustrate the usefulness of SGOOP-d while anticipating that the method directly applies to other biasing protocols as well.\cite{tiwary2016review}  We demonstrate the utility of the method by studying a range of model systems including conformational transitions in a small peptide Ace-Ala$_3$-Nme. In this system, for instance, one has a total of at least 28 inter-state transitions. As we show here, with only two component-RC learned from SGOOP-d we do accurately capture most of the 28 pairs of distances, with minimal improvement achieved by adding a 3rd component to the RC. Similar results are obtained on the basis of input trajectories coming from metadynamics simulations biased along pre-selected biasing variables. 
Open-source software detailing the method has also been released.

\begin{figure}[t!]
  \centering
  \includegraphics[width=8cm]{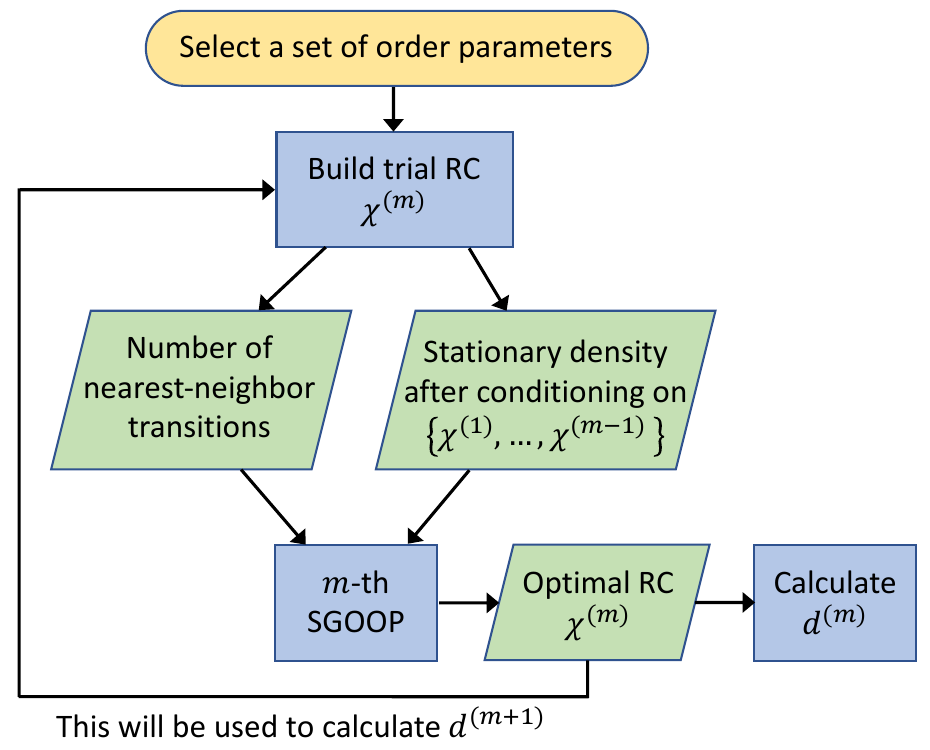}
  \caption
  { This flowchart describes the calculation of the $m$-th RC component $\chi^{(m)}, m \geq 1$ through multi-dimensional spectral gap optimization. For each $m$ we calculate $d^{(m)}$ in Eq.~\ref{eq:cd_2RCs}, which represents the contribution to commute distance on the basis of this $m$-th component. The optimal RC $\chi^{(m)}$ will be fed to the next SGOOP calculation for finding $d^{(m+1)}$. This iteration will stop when we obtain convergence in state-to-state $d^2_{\rm comm}$ values with addition of RC components. 
  The commute distance $d^2_{\rm comm}$ will be the sum of all the $d^{(m)}$ obtained in the iteration.
}\label{fig:flowchart}
\end{figure}

\section{Theory}
\label{sec:theory}
\subsection{Commute Distance and Commute Map}
\label{sec:commuted}
Our work builds upon the powerful advances first introduced by No\'{e}, Clementi, and co-workers that allow quantifying a kinetically truthful distance metric between generic molecular configurations.\cite{noe2015kinetic,noe2016commute} One such notion of ``kinetic distance" was introduced in Ref. \onlinecite{noe2015kinetic}, which was then generalized in Ref. \onlinecite{noe2016commute} as the ``commute distance". Both of these distances amount to transformations of the input coordinate space into a new space wherein Euclidean distances directly correspond to interconversion times. Here we summarize the basic ideas which originated from diffusion maps~\cite{nadler2006diffusion,coifman2005geometric} but were later generalized to Markovian dynamics.\cite{noe2015kinetic,noe2016commute} 

We consider a generic dynamical system undergoing Markovian dynamics in a finite-dimensional state space $\Omega$. The local density $\rho_t(\mathbf{x})$, $\forall\mathbf{x}\in\Omega$ can be propagated in time $t$ through
\begin{align}
\rho_{t+\tau}(\mathbf{y})=\int_{\mathbf{x}\in\Omega}\rho_{t}(\mathbf{x})p_{\tau}(\mathbf{y}|\mathbf{x})d\mathbf{x} \equiv \mathcal{P}\circ\rho_{t}(\mathbf{x})
\label{eq:propagator}
\end{align}
where $p_{\tau}(\mathbf{y}|\mathbf{x})$ is the transition density of finding the system at state $\mathbf{y}$ at time $t+\tau$ given that we have started it at state $\mathbf{x}$ at time $t$. Equivalently, Eq. \ref{eq:propagator} defines a Markov operator $\mathcal{P}$ and describes how an initial distribution $\rho_t(\mathbf{x})$ at time $t$ propagates to the distribution $\rho_{t+\tau}(\mathbf{y})$ at a later time $t+\tau$. One usual assumption made here is that there exists a unique equilibrium distribution $\pi(\mathbf{x})$ which satisfies
\begin{align}
\pi(\mathbf{x})=\mathcal{P}\circ\pi(\mathbf{x})
\label{eq:stationary_state}
\end{align}
At the same time, we can write an equivalent equation for the weighted density $\nu_t(\mathbf{x})=\rho_t(\mathbf{x})/\pi(\mathbf{x})$
\begin{align}
\pi(\mathbf{y})\nu_{t+\tau}(\mathbf{y})=\int p_{\tau}(\mathbf{y}|\mathbf{x})\pi(\mathbf{x})\nu_t(\mathbf{x})d\mathbf{x}=\mathcal{T}\circ\nu_t (\mathbf{x})
\label{eq:backward_propagator}
\end{align}
where $\mathcal{T}$ is the corresponding backward operator, also called the transfer operator. With this formalism, following the literature on diffusion maps\cite{coifman2005geometric} one defines a distance measure $D^2_{\tau}(\mathbf{x}_1, \mathbf{x}_2)$ between two points $\mathbf{x}_1, \mathbf{x}_2$ in the state space of a random walk as
\begin{align}
D^2_{\tau}(\mathbf{x}_1,\mathbf{x}_2)=\int_{\mathbf{y}\in\Omega}\frac{|p_{\tau}(\mathbf{y}|\mathbf{x}_1)-p_{\tau}(\mathbf{y}|\mathbf{x}_2)|^2}{\pi(\mathbf{y})}d\mathbf{y}
\label{eq:kd}
\end{align}
This definition can be seen\cite{coifman2005geometric} as equivalent to (a) preparing two ensembles initially located at $\mathbf{x}_1$ and $\mathbf{x}_2$, (b) letting them evolve by a lag time $\tau$, and then (c) computing the difference between the subsequently resulting probability distributions. In order to make use of Eq.~\ref{eq:kd}, one needs the transition density $p_{\tau}(\mathbf{y}|\mathbf{x})$. To facilitate its computation,\cite{noe2015kinetic} we assume that the transfer operator $\mathcal{T}$ has $N$ discrete eigenpairs and assume reversible dynamics/detailed balance $\pi(\mathbf{x})p_{\tau}(\mathbf{y}|\mathbf{x})=\pi(\mathbf{y})p_{\tau}(\mathbf{x}|\mathbf{y})$:
\begin{align}
&p_{\tau}(\mathbf{y}|\mathbf{x})=\sum^{N-1}_{j=0}\lambda_j(\tau)\psi_j(\mathbf{x})\pi(\mathbf{y})\psi_j(\mathbf{y})
\label{eq:eigendecomp}
\end{align}
where $\lambda_j$ and $\psi_j$ are the corresponding eigenvalues and eigenvectors of the transfer operator $\mathcal{T}$. With the orthonormality condition $\int\pi(\mathbf{y})\psi_j(\mathbf{y})\psi_k(\mathbf{y})d\mathbf{y}=\delta_{jk}$, applying Eq.~\ref{eq:eigendecomp} to Eq.~\ref{eq:kd} directly leads to:
\begin{align}
D^2_{\tau}(\mathbf{x}_1, \mathbf{x}_2)=\sum^{N-1}_{j=1}(\lambda_j\psi_j(\mathbf{x}_1)-\lambda_j\psi_j(\mathbf{x}_2))^2
\label{eq:kd_2}
\end{align}
In Eq. \ref{eq:kd_2} the summation starts at $j=1$ since the $j=0$ eigenvector for the transfer operator $\mathcal{T}$ is a constant in $\mathbf{x}-$space. By further integrating out the lag time $\tau$ in Eq.~\ref{eq:kd_2}, we can make Eq.~\ref{eq:kd_2} insensitive to the choice of the lag time, and in this way we arrive at the definition of the commute distance $d^2_{\rm comm}$:
\begin{align}
&d^2_{\rm comm}(\mathbf{x}_1, \mathbf{x}_2)=\int^{\infty}_{0}D^2_{\tau}(\mathbf{x}_1, \mathbf{x}_2)d\tau \nonumber \\ 
&=\sum^{N-1}_{j=1}\left(\sqrt{\frac{t_j}{2}}\psi_j(\mathbf{x}_1)-\sqrt{\frac{t_j}{2}}\psi_j(\mathbf{x}_2)\right)^2
\label{eq:cd}
\end{align}
where $t_j = -{\frac{\tau}{\ln\lambda_j}}$ is the relaxation timescale associated with $j$th eigenvector. Often one uses the rate $k_j={t_j}^{-1}$ instead of the timescale.\cite{perez2013identification} Eq.~\ref{eq:cd} now has a Euclidean distance form and a direct physical meaning: it is approximately the average time the system spends to commute between two states.\cite{noe2016commute} The  distance $d_{\rm comm}$ is thus called the ``commute distance", and the associated mapping
\begin{align}
\mathbf{x} \mapsto ( \sqrt{\frac{t_1}{2}}\psi_1, ..., \sqrt{\frac{t_{N-1}}{2}}\psi_{N-1})
\label{eq:cmap}
\end{align}
is called the ``commute map".  

Assuming that the dynamics in the $\mathbf{x}-$space is Markovian and fully sampled giving access to eigenvalues and eigenvectors of $\mathcal{T}$, we can then use Eq.~\ref{eq:cd} to calculate a Euclidean distance which approximates the commute time in the $\mathbf{x}-$space.  It is also worth pointing out that in Eq. \ref{eq:cd} the timescales follow $t_1 \geq t_2 \geq ... 0$, which implies that the commute distance increases monotonically with consideration of further eigenvectors of $\mathcal{T}$, and that there is an increasingly vanishing contribution from every additional eigenvector that we consider. 
If such a distance can be obtained through Eq.~\ref{eq:cd}, it is very useful for analyzing high-dimensional trajectories arising from well-sampled simulations as shown for instance in Ref. \onlinecite{noe2015kinetic,noe2016commute}. However many if not most real-world applications are characterized by rare events, wherein the system stays trapped in the part of the configuration space it was initiated from and rarely visits other regions. Adequate and reliable sampling of the underlying configuration space thus remains a longstanding challenge in computational chemistry and physics. This implies that the eigenvectors and eigenvalues needed to evaluate the various terms in Eq.~\ref{eq:cd} are simply not available or far from reliable. In fact, the dominant first few components of the commute map could even serve as biasing coordinates along which the sampling could be enhanced through methods such as umbrella sampling, metadynamics, or others. This brings out the inverse nature of the problem wherein constructing an accurate commute distance depends on sufficient sampling of the eigenvalues and eigenvectors of the transfer operator, but the sampling itself could benefit greatly from the knowledge of the commute map. 
 
\subsection{Calculating commute distances for rare events}
\label{sec:commuted_rare}

In this section, we develop a formalism for obtaining commute distances in poorly sampled rare-event systems where access to $\mathcal{T}$ and its eigenvectors/eigenvalues is not straightforward. The central idea is to perform biased sampling to accelerate the exploration of the configuration space. Here we use metadynamics as the biased sampling method, but the developed formalism should be more generically applicable. While this basic idea is simple, there are, however, at least two major, immediate difficulties when applying Eq.~\ref{eq:cd} with metadynamics or other similar enhanced sampling methods. First, the use of any sort of biasing corrupts the kinetics of the system, critical to calculating accurate eigenvalues and eigenvectors of the transfer operator $\mathcal{T}$. Second, the biasing itself needs access to the slow modes of the system, which are the dominant components of the commute map in Eq. \ref{eq:cmap}. In SGOOP, \pt{described in Sec. \ref{sec:sgoop} and \ref{sec:sgoop_details}}, we find these slow modes from the transfer operator of such a transition matrix but only look at its dynamics along a 1-d coordinate. We refer to these slow modes as the reaction coordinate (RC) for the system.\cite{ma2005automatic,bittracher2018data} As mentioned in Sec. \ref{sec:commuted} the different components $\psi_i$ of the commute map have a vanishing relevance to the calculation of the commute distance as $i \gg 1$, and thus one can stop after the first few dominant components and bias these components in any biasing method of choice. However, without knowing the commute map, it is hard to calculate the dimensionality and components of the RC which would then be biased.

\subsubsection{Spectral Gap Optimization of Order Parameters (SGOOP) for 1-dimensional RC}
\label{sec:sgoop}
In this sub-section we summarize the ``Spectral Gap Optimization of Order Parameters (SGOOP)" method for optimizing a multi-dimensional RC.\cite{tiwary2016spectral,tiwary2017predicting,smith2018multi} In later sections, we use SGOOP to develop an approach that circumvents both of the above-described challenges. Summarily, SGOOP in its original form is a method for obtaining a one-dimensional RC given static and dynamic information about a multi-dimensional system by combining this information in a maximum Caliber or path entropy framework.~\cite{presse2013principles,ghosh2020maximum} SGOOP constructs the RC as a combination of pre-selected candidate order parameters $(s_1, ..., s_d)$, which can be thought of as a set of basis functions using which we are trying to describe our problem. The dimensionality $d$ is kept high enough so that dynamics in the high-dimensional $s-$space is likely Markovian, needed for the formalism described in Sec. \ref{sec:commuted}. The central ideas behind SGOOP\cite{tiwary2016spectral} in its original form can be summarized as the following three points: \newline
(i) It uses a reweighting protocol\cite{reweighting_jpcb_2015} to estimate the equilibrium distribution $P_0(s_1, ..., s_d)$ from an initial metadynamics simulation performed by biasing some trial RC. \newline
(ii) In addition, it uses short unbiased MD simulations to obtain dynamical observables pertaining to the system. These observables could be the position-dependent diffusivity or more typically, the number of nearest-neighbor transitions along some binned trial RCs.\newline
(iii) By combining (i) and (ii) SGOOP constructs the transition rate matrices $K$ which can then be formulated as follows:
\begin{align}
K_{mn}=
\begin{cases}
-\Lambda\sqrt{\frac{\pi_n}{\pi_m}}, & \text{if } n\neq m \\
-\sum_{k\neq m} K_{mk}, & \text{if } n=m
\end{cases}
\label{eq:sgoop_rate0}
\end{align}
where $\pi\equiv P_0$ is the stationary probability along any putative, spatially discretized RC $\chi$ with $n$ denoting the grid index and $\Lambda$ is a dynamical observable. As mentioned in point (i), the stationary distribution can be obtained from a long unbiased simulation or from a biased simulation followed by an appropriate reweighting. The dynamical variable $\Lambda$, as discussed in point (ii), can be calculated by the number of nearest-neighbor transitions $\langle N\rangle$ defined as
\begin{align}
\langle N\rangle=\sum_{\text{(m,n)}\atop\forall|m-n|=1}\pi_mK_{mn}N_{mn}
\label{eq:NN}
\end{align}
where $N_{mn}=1\forall|m-n|=1$ and $0$ otherwise. Plugging Eq.~\ref{eq:sgoop_rate0} into Eq.~\ref{eq:NN} we obtain an estimate of $\Lambda$ as:
\begin{align}
\Lambda=\frac{\langle N\rangle}{\sum\sqrt{\pi_m\pi_n}}
\label{eq:Lambda}
\end{align}
The eigenvalues $\{k_j\}$ of the rate matrix $K$ are nonnegative and satisfy $k_0=0<k_1\leq k_2\leq ...$. The quantity $e^{-k_{n-1}}-e^{-k_{n}}$, which is the ``spectral gap" of the transfer operator $\mathcal{T}$, can be interpreted as the timescale separation between the $n$ slow mode and all the other hidden faster modes as projected on the corresponding RC. It can be shown that the optimal RC has the maximal spectral gap.\cite{tiwary2017predicting} Different candidate one-dimensional RCs are then first ranked in terms of the number of slow modes or metastable states they demarcate, and then in terms of the timescale separation (or the spectral gap) between the slow and fast modes as projected on any RC. The optimal RC maximizes both of these.

\subsubsection{SGOOP for multi-dimensional RCs and rate matrices}
\label{sec:sgoop_details}

We recently also introduced a multi-dimensional version of SGOOP \cite{smith2018multi} which makes it possible to extend the dimensionality of the RC in SGOOP. Each additional RC component $\chi^{(i)},i\geq 2$ is constructed in a way that it captures features indiscernible in the previous components through a conditional probability factorization described in Sec. \ref{sec:sgoop_details}. This de-emphasizes the features already captured by the components so identified. With multiple iterations of the SGOOP protocol one can identify a multi-dimensional RC $\chi = \{ \chi^{(1)},\chi^{(2)},...\}$. Mathematically this can be written as follows. Once the first RC component $\chi^{(1)}$ has been learned by SGOOP, we focus our attention on the probability distribution $P_1$ conditional on the knowledge of $\chi$ defined as:
\begin{align}
P_1(s_1, ..., s_d)&\equiv P_0(s_1, ..., s_d|\chi^{(1)}) \nonumber \\
&=\frac{P_0(s_1, ..., s_d)}{P_0(\chi^{(1)})}
\label{eq:CPF}
\end{align} 
where we have used that the equilibrium probability $P_0(s_1, ..., s_d,\chi^{(1)} )=P_0(s_1, ..., s_d)$ as $\chi^{(1)}$ is a deterministic function of $(s_1, ..., s_d)$. The next round of SGOOP is then performed on data sampled from $P_1$ instead of $P_0$, which yields the second RC component $\chi^{(2)}$ that captures features missed by $\chi^{(1)}$. The procedure can be repeated for further RC components and can be performed using any enhanced sampling method.\cite{smith2018multi} Here we illustrate it using metadynamics. By performing well-tempered metadynamics simulation along $\chi^{(1)}$ where one builds a bias $V_b (\chi^{(1)})$, it can be shown that
\begin{align}
P_0(\chi^{(1)})& \propto e^{-\beta F(\chi^{(1)})} \propto e^{+\beta[\frac{\gamma}{\gamma-1}V_b(\chi^{(1)})]} \nonumber \\
P_1 \equiv \frac{P_0(s_1, ..., s_d)}{P_0(\chi^{(1)})}& \propto e^{-\beta[F(s_1, ..., s_d)+V_b(\chi^{(1)})]}
\label{eq:P0P1}
\end{align}
where $\beta=1/k_B T$, $\gamma$ is the bias factor for well-tempered metadynamics,\cite{valsson2016enhancing} and $F$ is the free energy of the system. Therefore, $P_1$ is simply the unreweighted/biased probability density obtained by sampling in the presence of bias potential $V_b (\chi^{(1)})$.

We now discuss details of the construction of the rate matrix through SGOOP. 
Following \pt{Eq. \ref{eq:sgoop_rate0} and Eq. \ref{eq:NN}}, the rate matrix along any putative RC $\chi$ can be built as follows:
\begin{align}
K^{(1)}_{mn}=
\begin{cases}
-\frac{\langle N\rangle}{\sum\sqrt{\pi_n\pi_m}}\sqrt{\frac{\pi_n}{\pi_m}}, & \text{if } n\neq m \\
-\sum_{k\neq m} K_{mk}^{(1)}, & \text{if } n=m
\end{cases}
\label{eq:sgoop_rate1}
\end{align}
where $\langle N\rangle$ is the total number of nearest-neighbor transitions per unit time, counted along a suitably discretized RC $\chi = \{\chi_n\}$ with $n$ indicating grid index, $\pi \equiv P_0 $ is the corresponding stationary density \pt{and 1 in superscript indicates this is the rate matrix along the first component $\chi^{(1)}$ of the RC.} For the first round of SGOOP to learn $\chi^{(1)}$, $\langle N\rangle$ is calculated from short unbiased MD simulations. The $K^{(1)}$ matrices are then constructed for different putative RCs and its eigenvalues used to screen for the best RC $\chi^{(1)}$ with highest spectral gap.

For learning the second component $\chi^{(2)}$ and other higher-order components, we generalize Eq. \ref{eq:sgoop_rate1} as follows:\cite{smith2018multi}
 \begin{align}
K^{(2)}_{mn}=
\begin{cases}
-\frac{{\langle N\rangle}^{(1)}}{\sum\sqrt{\pi_n^{(1)}\pi_m^{(1)}}}\sqrt{\frac{\pi_n^{(1)}}{\pi_m^{(1)}}}, & \text{if } n\neq m \\
-\sum_{k\neq m} K_{mk}^{(2)}, & \text{if } n=m
\end{cases}
\label{eq:sgoop_rate2}
\end{align}
In Eq. \ref{eq:sgoop_rate2}, $\pi^{(1)} \equiv P_1 $ is defined in Eq. \ref{eq:P0P1}. $\langle N\rangle^{(1)}$ denotes the average number of first-nearest neighbor transitions along a putative RC observed per unit time, but now measured in the biased simulation performed by sampling from this conditional probability density $P_1$. The procedure can then be easily generalized for constructing rate matrices $K^{(3)},K^{(4)},...$ for learning further RC components.


\subsubsection{Commute distance calculation for rare events with SGOOP}
\label{sec:commute_with_sgoop}
Here we use SGOOP to induce a commute distance metric for complex high-dimensional systems that can be calculated from a combination of biased simulations and short  unbiased trajectories. Assuming that a satisfactorily large number of components have been included in $\chi$, any two points $\{ \mathbf{x},\mathbf{x'}\} \in\Omega$ can then be mapped without substantial loss  of information to its values in the $\chi$ space as $\{ \chi,\chi' \} $. Whether the dimensionality of the RC $\chi$ is indeed sufficient or not is a non-trivial question to answer, which we will address later in this section and in Sec. \ref{sec:results}. With the RC optimized by SGOOP, we can then reformulate Eq.~\ref{eq:cd} as  
\begin{align}
d^2_{\rm comm}(\mathbf{x},\mathbf{x'}) 
&= d^2_{\rm comm}(\chi, \chi') \nonumber \\
& =\sum^{N-1}_{j=1}\frac{1}{2k_j}\left[\psi_j(\chi)-\psi_j(\chi')\right]^2 \nonumber \\
& =\sum^{N-1}_{j=1}\frac{1}{2k_j^{(1)}}\left[\psi_j^{(1)}(\chi)-\psi_j^{(1)}(\chi')\right]^2 
\label{eq:d_comm2}
\end{align} 
In the above equation, we have made use of the mapping $\mathbf{x} \to \chi$ learned from SGOOP, but otherwise, it still needs the eigenvalues and eigenvectors of the transfer operator $\mathcal{T}$. In the final line, we have introduced a superscript $(1)$ to indicate the case where the first RC $\chi^{(1)}$ learned from SGOOP is indeed sufficient for the system at hand. In such a case, SGOOP yields a Maximum Caliber based rate matrix $K^{(1)}$ for transitions between grid points along  suitably discretized $\chi^{(1)}$. Details of the construction of this rate matrix are described in Sec. \ref{sec:sgoop_details} while illustrative examples are provided in Sec. \ref{sec:results}. By diagonalizing the rate matrix $K^{(1)}$ we obtain the eigenvalues $k_1^{(1)}, k_2^{(1)},...$ and corresponding eigenvectors $\psi_1^{(1)}, \psi_2^{(1)},...$ to use in Eq. \ref{eq:d_comm2}. 

The above commute distance so obtained can be understood as an estimate of true commute distance using the 1-dimensional projected RC $\chi^{(1)}$.  However, as shown in Sec. \ref{sec:results} and also emphasized in the literature on numerous occasions,\cite{altis2008construction} a 1-dimensional projection is often not kinetically truthful and does not reflect the connectivity of underlying high-dimensional space. We thus consider additional RC components $\chi^{(m)}$ from the multi-dimensional SGOOP protocol, with eigenvalues $k_1^{(m)}, k_2^{(m)},...$ and corresponding eigenvectors $\psi_1^{(m)}, \psi_2^{(m)},...$, where $m\geq 1$ denotes which RC component we are looking at. Each such component induces its own contribution to the commute distance which we add to the contribution of the 1st component $\chi^{(1)}$ in Eq. \ref{eq:d_comm2} yielding the central equation of this work for a $M-$component RC: 
\begin{align}
&d^2_{\rm{comm}}(\mathbf{x},\mathbf{x}') \nonumber \\
&=\sum^{M}_{m=1}\sum^{N-1}_{j=1}\frac{1}{2k^{(m)}_j}\left[\psi_j^{(m)}(\chi)-\psi_j^{(m)}(\chi')\right]^2 \equiv\sum^{M}_{m=1}d ^{(m)}
\label{eq:cd_2RCs}
\end{align}
Here $d ^{(m)}$ is the contribution to the commute distance arising from the $m^{th}$ RC component, while $k^{(m)}_j$ and $\psi^{(m)}_j$ are the $j^{th}$ eigenvalue and eigenvector of the Maximum Caliber-based transition matrix $K^{(m)}$ calculated along along RC-component $\chi^{(m)}$ (Sec. \ref{sec:sgoop_details}).

We want to mention two important points here. Firstly, for any RC component $\chi^{(m)}$ for $m\geq 1$, the construction of the rate matrix $K^{(m)}$ as detailed in Sec. \ref{sec:sgoop_details} ensures that the rates are ordered as per $0 < k_1^{(m)} \leq k_2^{(m)} \leq ... $. This leads to a useful property that the commute distance is a strictly monotonically increasing function of adding further RC components as well as further eigenvectors along any RC component. By monitoring how $d^2_{\rm{comm}} = d ^{(1)} + d ^{(2)}+...$ converges with addition of RC components, we can quantify the dimensionality of the RC needed for a given system at hand.  Secondly, the intuitive idea behind going from Eq. \ref{eq:d_comm2} to Eq. \ref{eq:cd_2RCs} is that different eigenvectors are orthogonal to each other allowing for a Euclidean distance measure. 
This is strictly true for the SGOOP-derived eigenvectors along a given RC component, i.e. the dot product of $\psi_j^{(m)}$ and $\psi_k^{(m)}$ is 0 $\forall j,k,m \geq 0$ as mentioned in Sec. \ref{sec:sgoop_details}. However when comparing $\psi_j^{(m)}(\chi^{(m)})$ and $\psi_k^{(n)}(\chi'^{(n)})$ for $m \neq n$ i.e. for different RC components through multiple rounds of SGOOP\cite{smith2018multi} this is not strictly true, and thus we expect Eq. \ref{eq:cd_2RCs} to be an upper bound for the commute distance. Note that the error could come from any eigenpair of each SGOOP rate matrix arising from redundant contributions due to different RC components having some aspects of the same dynamical processes. However as we will show later in Sec.~\ref{sec:results}, as long as each optimal RC captures the most important features or slowest processes, in the next round of SGOOP, such optimal RC will efficiently reduce the error from the non-orthogonality, making Eq.~\ref{eq:cd_2RCs} a good approximation to those important features.

\section{Results}
\label{sec:results}
In this section, we demonstrate the usefulness and reliability of the  SGOOP\cite{tiwary2016spectral,smith2018multi} based commute distance\cite{noe2015kinetic,noe2016commute} protocol developed in Sec. \ref{sec:theory}, which we label ``SGOOP-d" for convenience,  by applying it to a range of analytical potentials, as well as to small molecules with rare conformational transitions between different metastable states. Low-dimensional projections of these high-dimensional potentials can in general lead to a spurious number of barriers and inter-basin connectivity.\cite{altis2008construction,tsai2020learning} Here we show how to use SGOOP-d to ascertain the minimal dimensionality of the RC that preserves the kinetic aspects of the underlying high-dimensional landscape. To do so we calculate the state-to-state commute distances and monitoring how these change and eventually converge with an increase in RC dimensionality. This is done using either biased or long unbiased simulations. We can also use the RC so learned to perform further efficient and reliable biased simulations. We consider different types of unbiased and biased trajectories to demonstrate the general applicability of our proposed framework. Numerical and computational details of these systems have been provided in the Supplementary Information (SI). 

\begin{figure*}[t!]
  \centering
  \includegraphics[width=0.9\textwidth]{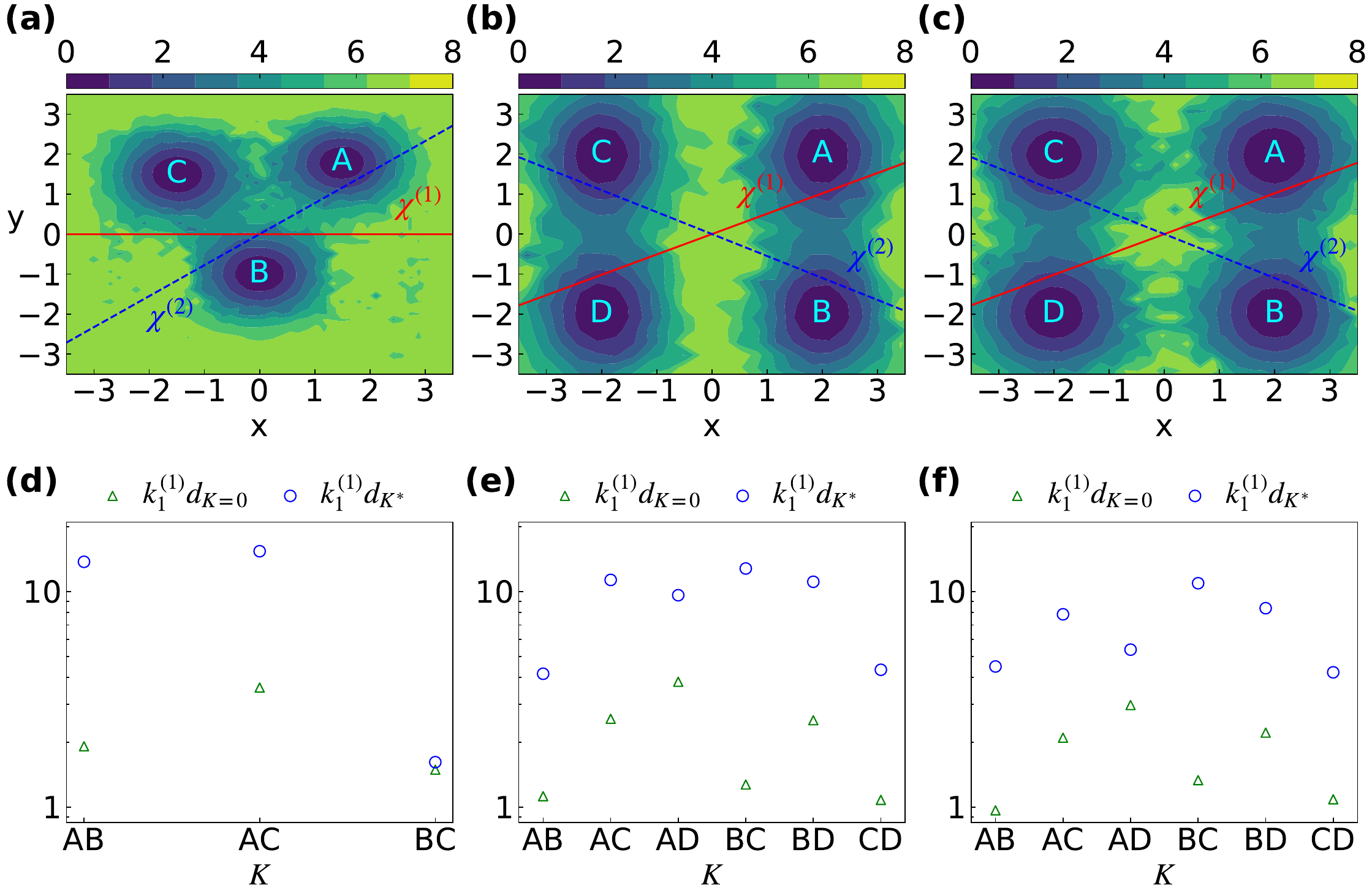}
  \caption
  { (a)-(c) show the 3-state and 4-state potentials 4A, 4B as sampled during molecular dynamics respectively. In (a)-(c) we have also provided the two RC components  $\chi^{(1)}$ (solid red lines) and $\chi^{(2)}$ (dashed blue lines) evaluated using Eq.~\ref{eq:dK}. Contours in all plots are separated by 0.89$k_B T$. In (d)-(f) we show the estimated commute distances $d_K$ between different pairs of metastable states (in arbitrary units) at $K=0$ and $K=K^{\ast}$. As explained in Sec. \ref{sec:analytical}, using $K=K^{\ast}$ gives the right kinetic connectivity between different metastable states for each of the model potentials. The results with statistical averages and error bars are shown in SI. Here we only show the result with one pair of RC for each model system in order to show how the second RC component captures the missing features of the first RC component.
}\label{fig:dK_3s4s}
\end{figure*}

\begin{figure*}[t!]
  \centering
  \includegraphics[width=0.95\textwidth]{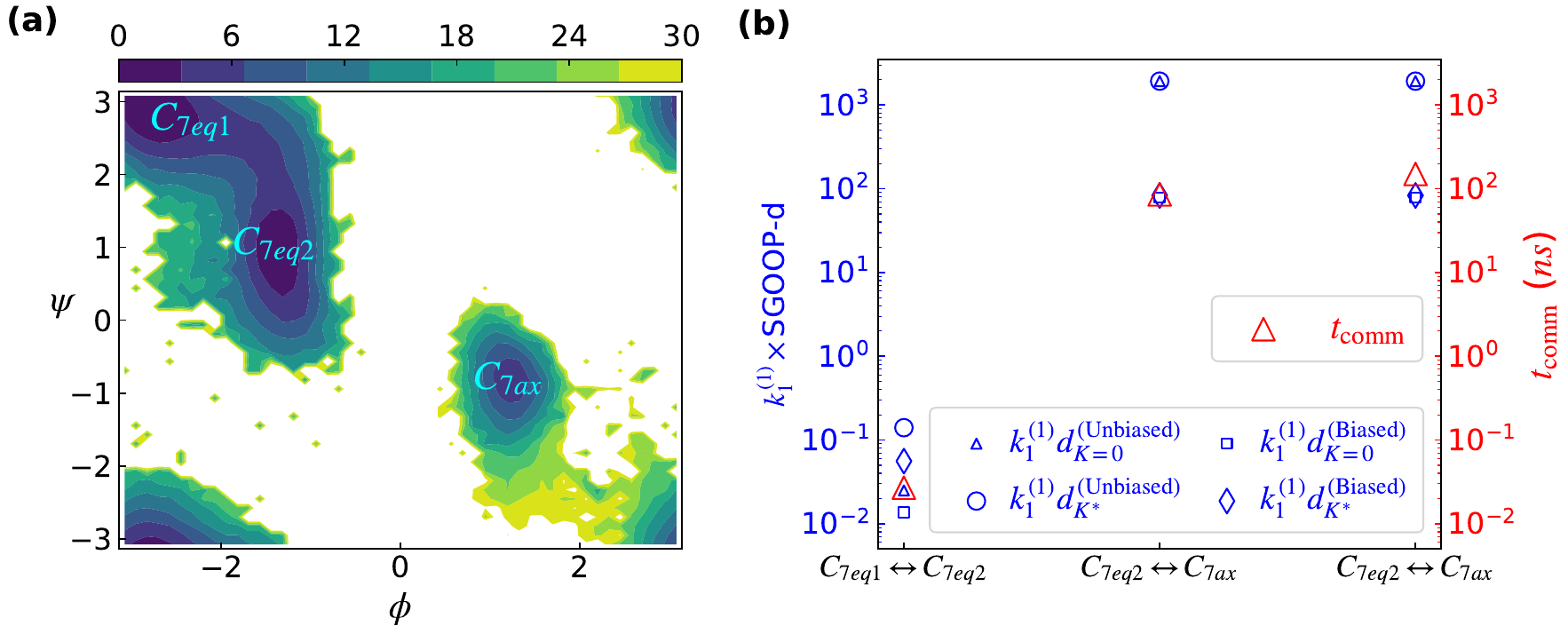}
  \caption
  {(a) Free energy surface as a function of $\phi$ and $\psi$ obtained by reweighting metadynamics simulation biasing along 1-d RC $\chi^{(1)}$ specified in Table. \ref{tab:RCala} . The positions of three metastable states are specified. (b) shows the SGOOP-d $k^{(1)}d_K$ at $K=0$ using one RC and at $K=K^{\ast}$ using two RCs for each pair of metastable states (in arbitrary units) obtained from a long unbiased simulation (blue triangles and blue circles respectively, left axis) and the biased simulation (blue squares, blue diamonds, left axis). In (b), we also provide the estimated commute time $t_{comm}$(red triangles, right axis) calculated from the long unbiased  simulation.
  }
  \label{fig:dK_aladip}
\end{figure*}

\begin{figure*}[t!]
  \centering
  \includegraphics[width=0.9\textwidth]{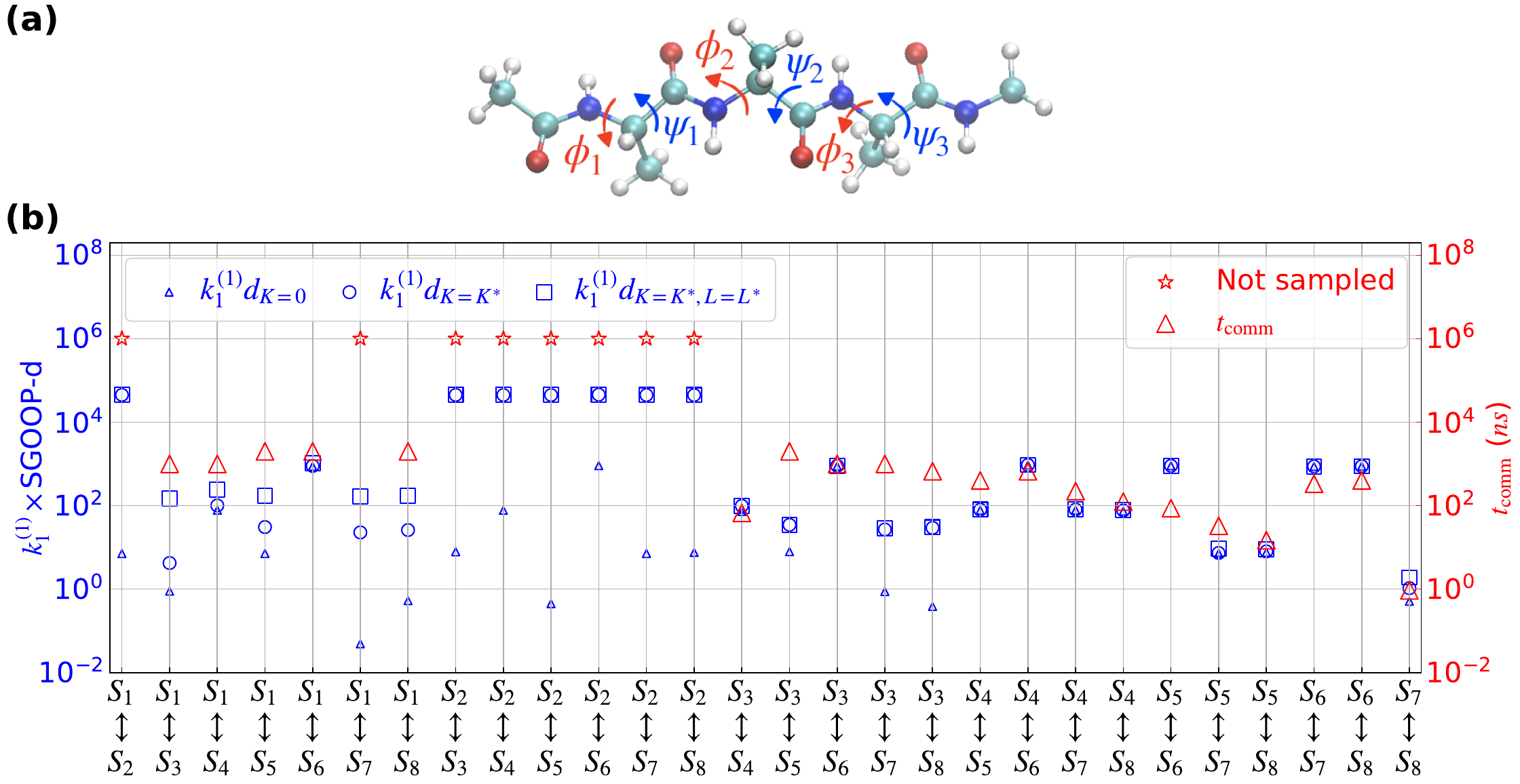}
  \caption
  {In this figure, (a) provides the molecular structure of Ace-Ala$_3$-Nme with the corresponding dihedral angles. The corresponding metastable states and their conformations are detailed in SI. (b) shows the calculation of SGOOP-d which provides the estimated commute distances using one-dimensional, two-dimensional and three-dimensional RC respectively (blue  triangles, blue circles and blue squares, left axis). The coefficients of these RCs are shown in Table~\ref{tab:RCala}. Corresponding to their calculation, these are labelled respectively $k^{(1)}_1d_{K=0}$, $k^ {(1)}_1d_{K^{\ast}}$ and $k^{(1)}_1d_{K^{\ast},L^{\ast}}$ (in arbitrary units) as shown in the legend.  (b) also provides the estimated commute time $t_{\rm comm}$ (red triangles, right axis) calculated from long unbiased simulation of Ace-Ala$_3$-Nme. The slowest transitions which are not sampled in the long unbiased simulation are denoted by star markers in the plot. Their commute times are not quantitatively reliable and serve only as guide to the eye.}
  \label{fig:dK_alatetra}
\end{figure*}

\subsection{Analytical potentials}
\label{sec:analytical}
 The analytical potentials used here are originally inspired from Ref. \onlinecite{altis2008construction}. These are built with two degrees of freedom $x$ and $y$, but with a varying number of metastable states and barriers separating them. Thus a 1-d projection is not always guaranteed to be kinetically truthful. Specifically we consider a 3-state potential and two 4-state potentials labeled 4A and 4B (Figs. \ref{fig:dK_3s4s} (a)-(c)). For each of these, we build inter-state commute distances using one-dimensional and two-dimensional RCs, with different components expressed as linear combinations of $x$ and $y$. Since the underlying dimensionality is two, here we will demonstrate the results with up to two-dimensional RC.
In such a case we can simplify Eq.~\ref{eq:cd_2RCs} by introducing
\begin{align}
\hat{d}^{(m)}&=k^{(m)}_1d^{(m)}
\label{eq:simplify_dK}
\end{align}
and then writing
\begin{align}
d_{\rm comm}(\mathbf{x}_1, \mathbf{x}_2)&=d^{(1)}+d^{(2)} \nonumber \\
&=\frac{1}{k^{(1)}_1}\hat{d}^{(1)}+\frac{1}{k^{(2)}_1}\hat{d}^{(2)}
\label{eq:cd_2RCs_1eval}
\end{align}


To see how good a job the RC components do at reconstructing the state-to-state connectivity, we further parameterize Eq.~\ref{eq:cd_2RCs_1eval} by \pt{introducing a} $K \equiv \frac{k^{(1)}_1}{k^{(2)}_1}$ for the ratio of eigenvalues, yielding
\begin{align}
k^{(1)}_1 d_{\rm comm}(\mathbf{x}_1, \mathbf{x}_2) \equiv k^{(1)}_1 d_K\equiv\hat{d}^{(1)}+K\hat{d}^{(2)}
\label{eq:dK}
\end{align}
We highlight here that in our framework $K$ is not a free parameter that needs to be tuned. Instead, it can be approximated on the basis of Maximum Caliber based rate matrices (Sec. \ref{sec:sgoop_details}) as:
\begin{align}
K^{\ast}\equiv\frac{k^{(1)}_1}{k^{(2)}_1}
\label{eq:K_ast}
\end{align}
\pt{where $K^{\ast}$ indicates a Maximum Caliber based estimation of $K$}. However, as the Maximum Caliber-based rate estimates are approximate and might depend on the choice of the dynamical constraints and quality of sampling,\cite{ghosh2020maximum} in SI we also show that the precise value of $K^{\ast}$ doesn't have a large effect on the connectivity.

Fig.~\ref{fig:dK_3s4s} and Table \ref{tab:RC3s4s} detail the two RC-components $\chi^{(1)}$ and $\chi^{(2)}$ so obtained for the different model potentials. Here using $K=0$ is equivalent to using only the first component $\chi^{(1)}$ to determine the commute distance, while increasing non-zero values of $K$ captures increasing contributions from the second component $\chi^{(2)}$ through Eq. \ref{eq:dK}. As can be seen for the 3-state system (Fig. \ref{fig:dK_3s4s} (d)), considering only the first component $\chi^{(1)}$ would lead to an erroneous conclusion that the pairs of states AB, AC, and BC are all kinetically equidistant. This is not consistent with the high-dimensional data sampled shown in Fig. \ref{fig:dK_3s4s} (a), where the barrier experienced between the states BC is much lower than for AB and AC. By adding the second component $\chi^{(2)}$ to the kinetic distance in Eq. \ref{eq:dK} using $K=K^{\ast}$, we recover this correct picture. Similar conclusions regarding kinetically truthful picture consistent with the data can be drawn for the remaining two 4-state potentials shown in Fig. \ref{fig:dK_3s4s}.  In both Fig.~\ref{fig:dK_3s4s} (e) and (f), using only the 1-d RC $\chi^{(1)}$, AB, BC, and CD are equally short, while AD is the slowest transition. This erroneous connectivity has been corrected after adding a second component of RC $\chi^{(2)}$, where AB and CD are equally shortest at $K=K^{\ast}$. Note that in both Fig.~\ref{fig:dK_3s4s} (e) and (f) AD is slightly lower which shows the noisy nature in the Maximum Caliber-based estimation of transition rates.

\begin{table}[]
    \centering
    \begin{tabular}{|c|c|c|c|}
    \hline
         \multicolumn{2}{|c|}{Systems} & $\theta^{(1)}/\pi$ & $\theta^{(2)}/\pi$ \\ \hline
         \multicolumn{2}{|c|}{3-state} & 0.00  & 0.21 \\ \hline
         \multirow{3}{*}{4-state}
         & 4A (Fig. \ref{fig:dK_3s4s}(b)) & 0.15 & 0.84  \\ \cline{2-4}
         & 4B (Fig. \ref{fig:dK_3s4s}(c)) & 0.15 & 0.84 \\ \hline
    \end{tabular}
    \caption{ In this table, we have shown the first and second components of the reaction coordinate $\chi^{(1)}$ and $\chi^{(2)}$  found for each model analytical potential. The angles $\theta^{(1)}$ and $\theta^{(2)}$ in the table define $\chi^{(i)}=\cos(\theta^{(i)})x+\sin(\theta^{(i)})y$.
    }
    \label{tab:RC3s4s}
\end{table}

\begin{table*}[]
    \centering
    \begin{tabular}{|c|c|c|}
    \hline
         Systems & RCs & Coefficients \\ \hline
         \multirow{2}{*}{Alanine dipeptide} 
         & $\chi^{(1)}$ & $(0.643,0.778,-0.133,-0.088,-0.221,-0.165)$ \\
         \cline{2-3}
         & $\chi^{(2)}$ & $(0.827,1.166,-0.120,0.578,0.013,0.240)$ \\ \hline
         \multirow{3}{*}{Ace-Ala$_3$-Nme} 
         & $\chi^{(1)}$ & $(0.187, -1.127, -0.228, -2.362, 0.230, 1.176)$ \\ 
         \cline{2-3}
         & $\chi^{(2)}$ & $(1.174, 0.738, 0.132, 0.716, 0.356, 2.827)$ \\ 
         \cline{2-3}
         & $\chi^{(3)}$ & $(-0.037, -0.839, 0.557,  1.454,  1.693, 1.624)$ \\ 
         \cline{2-3}
         \hline
    \end{tabular}
    \caption{ This table shows the reaction coordinates found for alanine dipeptide and Ace-Ala$_3$-Nme. For alanine dipeptide, two RC components both expressed as $\chi=a\cos\phi+b\sin\phi+c\cos\psi+d\sin\psi+e\cos\theta+f\sin\theta$ with their 6 respective coefficients are listed. For Ace-Ala$_3$-Nme, three RC components all expressed as $\chi=a\cos\phi_1+b\sin\phi_1+c\cos\phi_2+d\sin\phi_2+e\cos\phi_3+f\sin\phi_3$ with their 6 respective coefficients are listed.
    }
    \label{tab:RCala}
\end{table*}

\subsection{Alanine dipeptide}
\label{sec:aladi}
The next system we use to illustrate our method is the well-studied alanine dipeptide. Here we consider the molecule as characterized by three dihedral angles $\phi$,$\psi$, and $\theta$. This molecule has three metastable configurations (Fig. \ref{fig:dK_aladip}(a)) which can be characterized by using only $\phi$ and $\psi$, while $\theta$ plays a role in characterizing the transition between the metastable states.\cite{tiwary2013metadynamics} Here we express the different RC components as linear combinations of 6 order parameters, namely cosines and sines of the 3 aforementioned dihedrals, with the final optimized coefficients listed in Table. \ref{tab:RCala}. The spectral gap in SGOOP is optimized using a basin-hopping algorithm.~\cite{wales2003energy,wales1997global,li1987monte} These RC components and associated information are then plugged into Eq. \ref{eq:dK} to estimate the commute distance $d_K$. In Figs. \ref{fig:dK_aladip}(b)-(c) we show the commute distance so calculated using an input biased trajectory and a benchmark long unbiased trajectory respectively. The biased trajectory was generated by doing well-tempered metadynamics along 1-d RC $\chi^{(1)}$ defined in Table. \ref{tab:RCala}. See SI for further details of both the biased and unbiased simulations.

For this simple system, the commute distances $d_K$ show similar connectivities for $K=0$ and $K=K^{\ast}$, which shows that one RC is indeed sufficient to describe the system in terms of recovering state-to-state connectivity between all 3 metastable states. Both types of input trajectories show a near degenerate structure with two pairs of states kinetically separated from each other, while one pair is very close.

\subsection{Ace-Ala$_3$-Nme}
\label{sec:alatri}
 In this final section, we demonstrate our method on a more complicated molecular system, namely the peptide Ace-Ala$_3$-Nme with a much larger number of metastable states, and an even larger number of state-to-state transitions.~\cite{rydzewski2020multiscale} Simulation details are provided in SI. As discussed in Ref.~\onlinecite{rydzewski2020multiscale} the three dihedral angles $\phi_1$, $\phi_2$, $\phi_3$  are sufficient to characterize the $2^3=8$ dominant metastable states corresponding to positive and negative parts of the Ramachandran diagram for the 3 central Alanine residues. The RC components used in computing SGOOP-d distances are calculated as a linear combination of cosines and sines of these 3 dihedral angles, thereby amounting to a total of 6 order parameters. We consider the 8 most dominant metastable states labelled $S_1$,..., $S_8$ and the associated $\binom{8}{2} = 28$ inter-state transitions. The corresponding dihedral angles for these 8 states are tabulated in the SI. Here we consider up to three RC components and demonstrate that after considering 3 components the commute distances converge especially for the slower state-to-state transitions.  They are also in agreement with the benchmark calculations on this system through counting transitions in the higher dimensional underlying space from a long unbiased trajectory.  The final optimized solutions for all three RC components are shown in Table.~\ref{tab:RCala}. Here in order to add a third RC component, we generalize Eq.~\ref{eq:dK} by introducing an additional parameter $L$:
\begin{align}
    k^{(1)}_1 d_{K,L}\equiv\hat{d}^{(1)}+K\hat{d}^{(2)}+L\hat{d}^{(3)}
\label{eq:dK3}
\end{align}
Similar to what was done for $K^{\ast}$ in Eq. \ref{eq:K_ast} we can approximate $L^{\ast}$ as
\begin{align}
    L^{\ast}\equiv\frac{k^{(1)}_1}{k^{(3)}_1}
\label{eq:L_ast}
\end{align}
With a long enough unbiased MD trajectory, we can also calculate the commute time $t_{\rm comm}$ between two metastable states through a simple counting protocol (see SI and Ref.~\onlinecite{tsai2020learning}). In Fig.~\ref{fig:dK_alatetra}, we show SGOOP-d distances calculated using Eq. \ref{eq:dK3} with 1, 2, and 3 RC components, and compare them with the corresponding 28 $t_{\rm comm}$ values between the 8 metastable states in the same plot. It can be seen from the plot that with only the use of two RC components SGOOP-d already provides converged estimates of relative inter-state connectivity and commute distances between 23 of the 28 pairs of states based on the visualization of 3-d free energy provided in SI. Here we must point out that there are eight transitions that are not sampled by even the reference long unbiased simulation, although SGOOP-d of those transitions clearly converged. Therefore, the comparison of SGOOP-d with respect to the unobserved transitions may need a more cautious evaluation instead of merely looking at the free energy. However, in order to get the correct connectivity for the remaining 5 pairs of states as well, we have to include the third RC component. We emphasize that in Fig. \ref{fig:dK_alatetra} the slowest 8 transitions have been given the same reference commute time for the sake of clarity, as we were unable to observe any such transition events even in the 1 $\mu$s long unbiased simulation. Thus the reference commute times for these states serve as approximate lower bounds to the true values and are denoted by star markers in the plot.


\section{Conclusion}
\label{sec:conclusion}
In summary, in this work we have developed a computationally efficient formalism labeled ``SGOOP-d" and summarized in the flowchart in Fig. \ref{fig:flowchart}, that can help towards solving a longstanding important problem in chemical physics and physical chemistry. Namely, how many dimensions should a projection from high-dimensions into low-dimensional reaction coordinates (RC) have, so that (1) the projection is kinetically and thermodynamically truthful to the underlying landscape, and (2) these minimal number of components can then be used to perform biasing simulations without fear of missing slow degrees of freedom. The formalism here makes the best of two different approaches, namely commute map \cite{noe2016commute} and SGOOP. \cite{tiwary2016spectral} This way it induces a distance metric which we call SGOOP-d that is applicable to biased rare event systems as well as unbiased trajectories with arbitrary quality of sampling. The kinetically truthful RC learned here can then also be used to improve the sampling quality of the biased simulation itself\cite{bussi2020using} or as a progress coordinate in path-based sampling methods.\cite{zuckerman2017weighted,elber2020milestoning,votapka2017seekr,jiang2018forward,defever2019contour} We thus believe that going forward our work represents a useful tool in the study of kinetics in rare event systems with multiple states and interconnecting pathways. 

\section{Supporting Information}
The Supporting Information contains: (1) Simulation details including the setup of model potential and simulations for alanine dipeptide and Ace-Ala$_3$-Nme; (2)A plot with analytical model potential and averaged SGOOP-d at $K=0$ and $K=K^{\ast}$ with errorbars and a table of RC corresponding to averaged SGOOP-d; (3) A plot showing the linear combination of SGOOP-d at different K; (4) Free energy plots of Ace-Ala$_3$-Nme; (5) A list of landmark dihedral angles for calculating Fig.~\ref{fig:dK_alatetra}; (6) Metadynamics parameters; (7) PLUMED code for computing dihedral angles.

\section{Acknowledgments}
\label{sec:acknowledgements}
The authors thank Yihang Wang, Dedi Wang, Luke Evans, Shashank Pant, En-Jui Kuo, and Yixu Wang for discussions. This work was supported by the National Science Foundation, Grant No. CHE-2044165. ZS was supported by University of Maryland COMBINE program NSF award DGE-1632976. Acknowledgment is made to the Donors of the American Chemical Society Petroleum Research Fund for partial support of this research (PRF 60512-DNI6). We also thank Deepthought2, MARCC, and XSEDE (projects CHE180007P and CHE180027P) for the computational resources used in this work. SGOOP and SGOOP-d code are available at \url{github.com/tiwarylab}.

\bibliographystyle{unsrtnat}
\bibliography{main}

\end{document}